\documentclass[12pt]{iopart}

\usepackage{graphicx}
\usepackage{dcolumn}
\usepackage{bm}
\usepackage{amssymb}
\usepackage{epsf}
\usepackage{color}
\usepackage[utf8]{inputenc}

\begin{document}

\title[]{Understanding the magnetic response of the Quantum Spin Liquid compound 1$T$-TaS$_2$ \\}

\author{Sudip Pal and S. B. Roy}
\address{
 UGC DAE Consortium for Scientific Research\\
 University Campus, Khandwa Road\\
 Indore-452001, India 
}

\ead{sudip.pal111@gmail.com,sindhunilbroy@gmail.com}
\date{\today}

\begin{abstract}
1$T$-tantalum disulphide (1$T$-TaS$_2$) is a promising material to realize quantum spin liquid state and fractional excitations. Here, we take a closer look on the temperature dependent magnetic susceptibility of 1$T$-TaS$_2$ at low magnetic fields to understand the formation of local magnetic moment in this system and its response as a function of temperature and field. We found that the susceptibility increases with reduction in the temperature even in the nearly commensurate charge density wave (CDW) phase, which exhibits metallic conductivity and hence is expected to exhibit temperature independent Pauli paramagnetic susceptibility. Therefore, it indicates that local moment starts forming in the nearly commensurate  CDW phase itself, which is well above the temperature where the evidence of Mott insulating state is found in resistivity measurement. In the commensurate CDW phase, temperature dependence of susceptibility significantly deviates from the Curie-Weiss law and does not show the magnetic properties expected from S = 1/2 spin on a two dimensional triangular lattice. Below $T$ = 10 K, susceptibility follows a power law dependence on the temperature, which indicates towards the formation of random singlet state in 1$T$-TaS$_2$.
\end{abstract}      
                       
\section{Introduction} 
In last few years, the interest on a new kind of magnetic state called quantum spin liquid (QSL) has increased significantly. The idea of QSL was first proposed by P. W. Anderson in 1$T$-TaS$_2$ in the context of the magnetic ground state of spin, S = 1/2 on a triangular lattice {\color{blue}\cite{Anderson1973,Fazekas1974}}. Later, it was found that the ground state of a Heisenberg spin system on a triangular lattice is actually the 120$^{\circ}$ antiferromagnetic state and additional ingredients such as frustration arising from next near neighbor coupling {\color{blue}\cite{manual}}, spatially anisotropic interaction {\color{blue}\cite{trumper}}, ring exchange interaction  {\color{blue}\cite{olex}} etc. may be necessary to have a spin liquid ground state {\color{blue}\cite{Bernu1994}}. At present, several inorganic materials, such as YbMgGaO$_4$, $\alpha$-RuCl$_3$, and organic compounds like $\kappa$-(BEDT-TTF)$_2$Cu$_2$(CN)$_3$, EtMe$_3$Sb[Pd(dmit)$_2$]$_2$ etc. have been found to show QSL behavior {\color{blue}\cite{Reviews1, Reviews2, Paddison2017,Banerjee2018,Pustogow2018,Miksch2021}}.

1$T$-TaS$_2$, in this respect, is a rare system which belongs to di-chalcogenide family and shows series of charge density wave (CDW) transitions and a Mott insulating state at low temperature. It is a normal metal above $T$ = 550 K and undergoes a transition to incommensurate CDW state, followed by transitions to nearly commensurate and commensurate CDW states, when it is cooled below $T$ = 350 and 180 K respectively {\color{blue}\cite{Thomson1994}}. However, the system continues to show metallic behavior in the incommensurate and nearly commensurate CDW states. In the commensurate CDW state, 13 Ta atoms form a cluster, known as star-of-David where twelve Ta atoms in a layer move towards a thirteenth central Ta atom, and a single cluster carries a localized moment of S = $\frac{1}{2}$ and forms a triangular lattice in the a-b plane {\color{blue}\cite{Yamamoto1983,LawLee2017}}. The magnetic state of 1$T$-TaS$_2$ is now being extensively investigated for several reasons. It raises some of the important fundamental issues, such as the formation of magnetic moment in a metallic system undergoing CDW transitions {\color{blue}\cite{Zhang2014}} and the behavior of the localized moments situated on a triangular superlattice of  star-of-David clusters with temperature and magnetic field.  Interestingly, 1$T$-TaS$_2$ exhibits a sharp rise in the magnetic susceptibility and resistivity at low temperatures, which further raised interest in this system over the time {\color{blue}\cite{DiSalvo1980,Kratochvilova2017,Sudip2019}}. There are different proposals to account for such transport and magnetic behavior {\color{blue}\cite{DiSalvo1980, Fazekas1980}}. Among them, the widely accepted one is that in the commensurate CDW state, it isolates a narrow band near the Fermi level, which contains 5$d$ orbital of the central Ta atom out of 13 Ta atoms star-of-David cluster. Therefore, moderate electron-electron correlation is able to open Mott-Hubbard gap in the conduction band and induce a transition to the Mott insulating state {\color{blue}\cite{Pillo2000}}. As a result, the resistivity starts increasing at low temperatures, and simultaneously, around the same temperature, magnetic susceptibility also increases rapidly down to the lowest temperature. However, a few recent experimental and theoretical evidences also suggest that the insulating state in 1$T$-TaS$_2$ may arise from inter layer locking due to stacking order along the c-axis {\color{blue}\cite{Band}}. Nonetheless, instead of undergoing a transition to long range magnetic order or spin glass state, the localized moments in the insulating state form a nontrivial quantum spin liquid state with spinon excitation as suggested by Law and Lee {\color{blue}\cite{LawLee2017,He2018}}. However, there exist some indication of the formation of antiferromagnetic ordering at very low temperature as well {\color{blue}\cite{Perfetti2005}}.

Neutron scattering and muon spin relaxation studies on 1$T$-TaS$_2$ have revealed an interesting quantum state with short range spin correlation at low temperatures {\color{blue}\cite{Kratochvilova2017,Ribak2017,Klanjsek 2017}}. In bulk magnetic measurements, the susceptibility of 1$T$-TaS$_2$ is nearly temperature independent at high temperatures and increases rapidly below T = 50 K. A modified Curie-Weiss (CW) law given by: $\chi = \chi_0 + \frac {C}{(T-T_0)}$  has been widely used to analyze such temperature dependence, where $\chi$ is the experimentally obtained susceptibility, C is the Curie constant, $T_0$ is the Curie temperature and $\chi_0$ is a fitting parameter which accounts for the temperature independent contribution to the susceptibility {\color{blue}\cite{Kratochvilova2017,Ribak2017}}. In general, Larmour diamagnetism, Pauli paramagnetism and Van Vlack paramagnetism can contribute to $\chi_0$ {\color{blue}\cite{Ashcroft}}. Such CW law is generally expected from a collection of spins with negligible exchange interaction. However, even in canonical spin-glasses like AuMn and AgMn a deviation from the CW law has been observed in the paramagnetic regime well above the spin-glass transition temperature, thus indicating the presence of short range magnetic correlation {\color{blue}\cite{majumdar1983}}. In the light of some evidences of short range correlation at low temperatures in 1$T$-TaS$_2$ as revealed in Neutron scattering and muon spin relaxation studies {\color{blue}\cite{Kratochvilova2017,Ribak2017,Klanjsek 2017}},  it is thus prudent to look into the temperature dependence of the susceptibility of 1$T$-TaS$_2$ more carefully. In the resonating valance bond picture of quantum spin liquid, two spins supposedly form a spin singlet and the entire system fluctuates between extensive degenerate states. However, quenched random disorder may drive the system to form a new state called as random singlet state. In addition, the system can have a fraction of unpaired spins known as orphan spins, which will affect the overall magnetic response of the system

It may also be noted here that an unexpected rise in the susceptibility at low temperatures, which is commonly termed as the 'Curie tail' has been observed in various systems, including a few QSL compounds, topological materials etc {\color{blue}\cite{Bordelon2019,Saleheen2020,Bowers1965,sudip2022}}. This 'Curie tail' is generally assumed to have an extraneous origin, such as lattice defects, paramagnetic impurities etc. However, in a recent study on high purity elemental copper metal we have shown that such 'Curie tail' can have an intrinsic origin {\color{blue}\cite{sudip2022}}. This finding has given us further impetus to have a relook into the bulk magnetic reposnse 1$T$-TaS$_2$. Here we will address in details to the temperature dependence of magnetic susceptibility in 1$T$-TaS$_2$ down T = 2 K and try to rationalize our findings on the basis of available theroretical and experimental information.
\begin{figure}[t]
\centering
\includegraphics[scale=0.40]{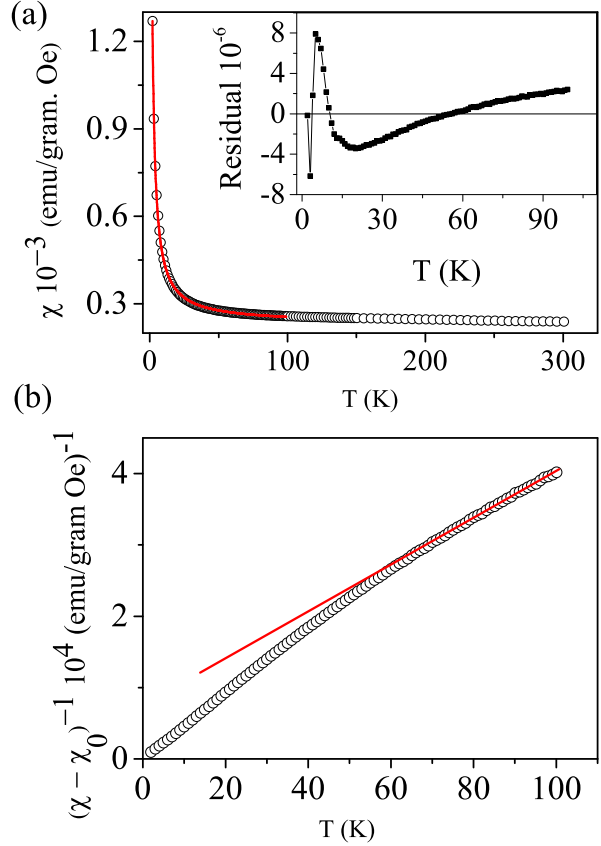}
\caption{(a) Experimentally obtained susceptibility ($\chi$) versus temperature ($T$) plot of  1$T$-TaS$_2$ at H = 1000 Oe. The red line below T = 100 K is the fitted curve using modified CW law. Inset shows the residual of the fitting. (b) $(\chi - \chi_0)^{-1}-T$ plot in the temperature range of 2 to 100 K at the measuring field of H = 1000 Oe. The red line indicates deviation from linearity.}
\end{figure}

\begin{figure}[t]
\centering
\includegraphics[scale=0.40]{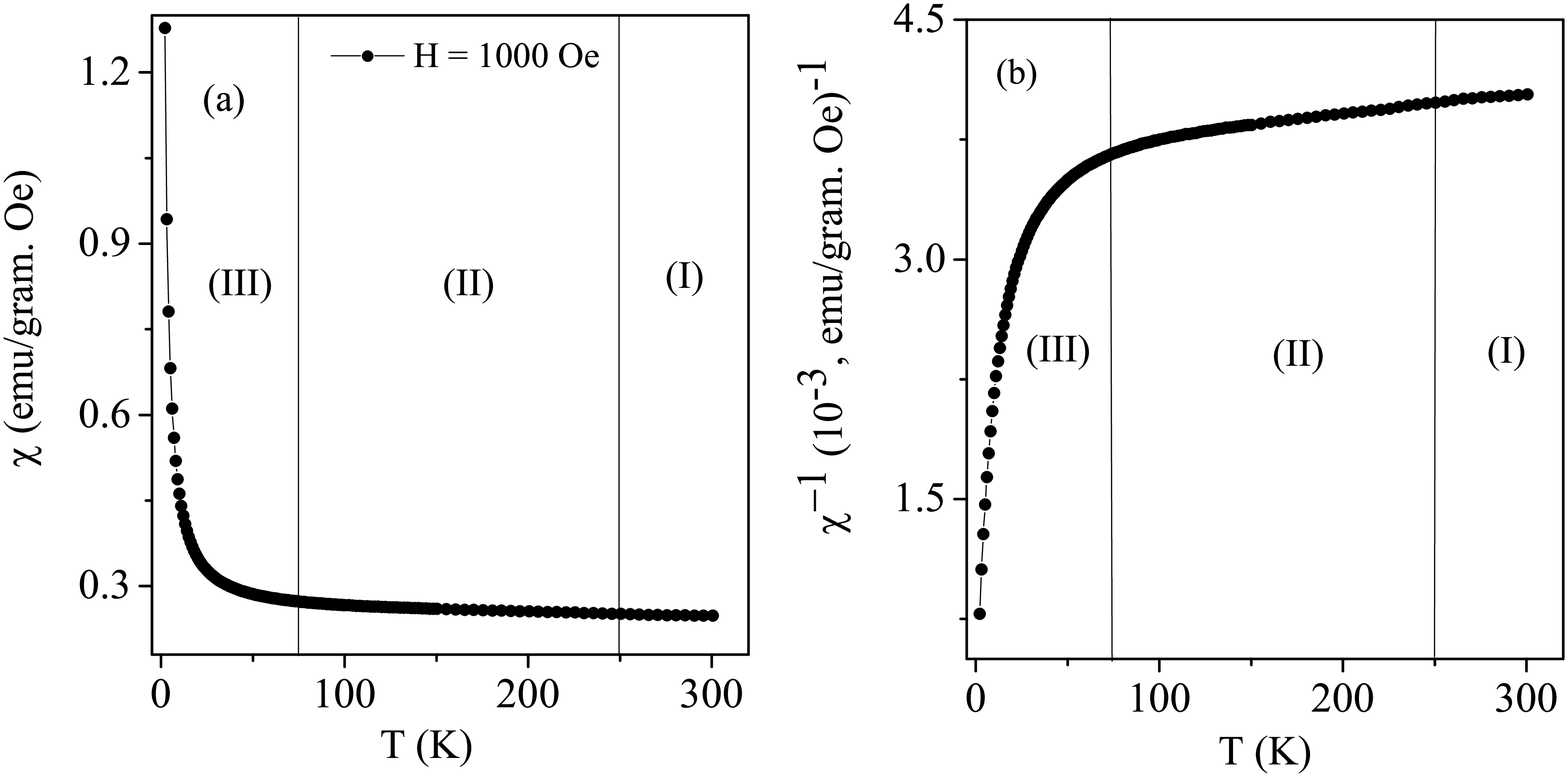}
\caption{(a) Susceptibility ($\chi$) versus temperature ($T$) and (b) inverse susceptibility ($\chi^{-1}$)  versus temperature ($T$) plots of  1$T$-TaS$_2$ in the temperature range of 2 to 300 K at the measuring field of H = 1000 Oe. We have divided the entire temperature region into three segments.}
\end{figure}

\section{Experimental details}
The single crystal sample has been prepared by chemical vapor transport technique. The details of the sample preparation can be found in earlier reports {\color{blue}\cite{Kratochvilova2017}}. Samples from the same batch have been used in earlier studies {\color{blue}\cite{Kratochvilova2017,Sudip2019}}. DC susceptibility measurements have been carried out in 7 Tesla SQUID magnetometer (M/S Quantum Design, USA). Temperature dependent dc susceptibility data has been recorded at $H$ = 1000 and 500 Oe between $T$ = 300 to 2 K in the zero field cooled (ZFC), field cooled cooling (FCC) and field cooled warming (FCW) protocols. In the ZFC protocol, the sample has been cooled down to $T$ = 2 K in absence of external field, and temperature dependence of susceptibility has been recorded during warming after the application of the measuring field.  Susceptibility has been measured in the subsequent cooling and heating cycles at the same field to record the FCC and FCW curves respectively. The temperature dependence of susceptibility at $H$ = 500 Oe has been reported earlier {\color{blue}\cite{Sudip2019}}. At this field, the FCC and FCW susceptibility curves are identical and the ZFC and FCW susceptibility curves show thermomagnetic irreversibility. Therefore, here we shall look into both the ZFC and FCW susceptibility curves at $H$ = 500 Oe. On the other hand, at $H$ = 1000 Oe, all three curves are merged in the probing temperature range and hence, only the ZFC curve will be presented. The temperature sweep rate during all the measurements has been kept at 1 K/min. It may be noted that to confirm that the signal is intrinsic to the sample under investigation, we have performed magnetic measurement of the sample holder. The magnetic response of the plastic sample holder has been found to be within the resolution of the system and it does not have any feature, such as upturn at low temperature as observed in 1$T$-TaS$_2$. This confirms that the observed magnetic response is not an instrumental artifact, but intrinsic to the 1$T$-TaS$_2$ material under investigation.
\section{Results and Discussion}
In Fig. {\color{blue}1(a)}, we have shown the temperature dependence of susceptibility of 1$T$-TaS$_2$ at the measurement field of H = 1000 Oe. As we cool the sample from room temperature, its susceptibility increases very slowly and below T = 100 K the susceptibility start increasing rapidly. Following earlier reports {\color{blue}\cite{Kratochvilova2017,Ribak2017}}, we have fitted the susceptibility below T = 100 K using a modified CW law given by $\chi = \chi_0 + \frac {C}{(T-T_0)}$. The fitted curve has been shown in Fig. {\color{blue}1(a)} (Red line). We want to highlight here that this fitting is not very satisfactory. In the inset, we have shown the residual of the fitting, which should have a random distribution in case of good match between experimental and calculated values. However, in this case, the difference between the fitted curve and the experimental values is not arbitrary in the fitted temperature range, which indicates that the modified CW law does not fit the experimental curve very well. In addition, the plot of $(\chi - \chi_0)^{-1}$ versus T is not linear within T = 2-100 K as we have shown in Fig. {\color{blue}1(b)}. The obtained value of the fitting parameter $\chi_0$=  2.33$\times$ $10^{-4}$ emu/gram Oe, which should arise from Larmour diamagnetic, Pauli paramagnetic and van Vleck paramagnetic contributions.  Importantly, we can estimate these temperature independent susceptibility contributions in 1$T$-TaS$_2$ using standard equations {\color{blue}\cite{Ashcroft,blundell}}. We show below that the value of the $\chi_0$ obtained from the fitting does not match with the theoretically estimated values. 

In 1$T$-TaS$_2$, the fully occupied electron shells will contribute a negative susceptibility due to Larmour diamagnetism. In addition, 1$T$-TaS$_2$ exhibits metallicity of some kind until Mott metal-insulator transition occurs in the commensurate CDW phase. Hence, a positive Pauli susceptibility due to itinerant electrons is expected at high temperatures. The diamagnetic susceptibility can be estimated following the text book formula {\color{blue}\cite{Ashcroft}}:
\begin{equation}
\chi_D = - 0.79Z_i/M \times 10^{-6} \langle (r/a_0) \rangle^2 ~emu/gram Oe
\end{equation}
Here, $Z_i$ is the number of electrons in the atom, $r$ is the atomic radius and $a_0$ is the Bohr radius. For 1$T$-TaS$_2$, diamagnetic susceptibility  is approximately $\chi_D = -1.65\times 10^{-6}$ emu/gram Oe.
On the other hand, the Pauli paramagnetic susceptibility at T = 0 K can be written as,
\begin{equation}
\chi_P = \mu_B^2 \times g(E_F)   
\end{equation}
where, $g(E_F)$ is the total density of states at the Fermi energy. Using a value of $g(E_F)$ = 1.37 eV$^{-1}$ {\color{blue}\cite{DOS}} we estimate a value of Pauli paramagnetic susceptibility for 1$T$-TaS$_2$  to be $\chi_P = 0.18\times 10^{-6}$ emu/gram Oe . This value of Pauli paramagnetic susceptibility is much smaller than the the value of $\chi_0$ obtained from the fitting of modified CW law. Furthermore, $\chi_P$ should vanish in the Mott insulator state of 1$T$-TaS$_2$ at low temperatures. In addition, the first order nature of the transition between nearly commensurate and commensurate CDW phases results into phase coexistence metal and CDW state in a wide range of temperature. Hence Pauli paramagnetism should also vary with the volume fraction of the metallic phase, and progressively decrease with the decrease in temperature even within the incommensurate and nearly commensurate CDW phases. Overall it is expected that Pauli paramagnetism is not going to make any meaningful contribution in the low temperature regime below 50 K.  Van Vleck paramagnetism arises due to change in the ground state energy with the application of magnetic field within the framework of second-order perturbation theory by taking into account a mixing with excited states with total non zero angular momentum  {\color{blue}\cite{blundell}}. However, in the temperature range below 50 K the possibility of such mixing with the excited state is small, and hence in this temperature range Van Vleck contribution to the susceptibility is negligible {\color{blue}\cite{Bordelon2019}} 

Therefore to understand the magnetic state of 1$T$-TaS$_2$ at low temperatures below 50 K,  it is necessary to focus on the net magnetic response from the unpaired local moments as a function of temperature. Henceforth, we will study the resultant magnetic susceptibility, $\chi = (\chi_{exp}- \chi_D$), where we have subtracted only the diamagnetic component $\chi_D$ from the experimentally obtained susceptibility $\chi_{exp}$. Fig. {\color {blue}2(a)} and {\color {blue}2(b)} show the temperature variation of $\chi$ and $\chi^{-1}$ respectively at $H$ = 1000 Oe in the temperature range from 2 to 300 K.

We can divide the entire temperature range into a few segments where the susceptibility shows distinct behavior as temperature is reduced from room temperature: ($i$) $T>250$ K, , M shows slow but noticeable linear increase with decrease in the temperature, ($ii$) $100< T<250$ K where susceptibility shows gradual but relatively fast increase in the susceptibility, and  ($iii$) $T<100$ K, where magnetization exhibits a sharp rise down to lowest temperature of measurement, which in our case is 2 K. These features are more clearly visible in the temperature dependence of inverse susceptibility data as shown in Fig. {\color{blue}2(b)}.\\  
\begin{figure}[t]
\centering
\includegraphics[scale=0.38]{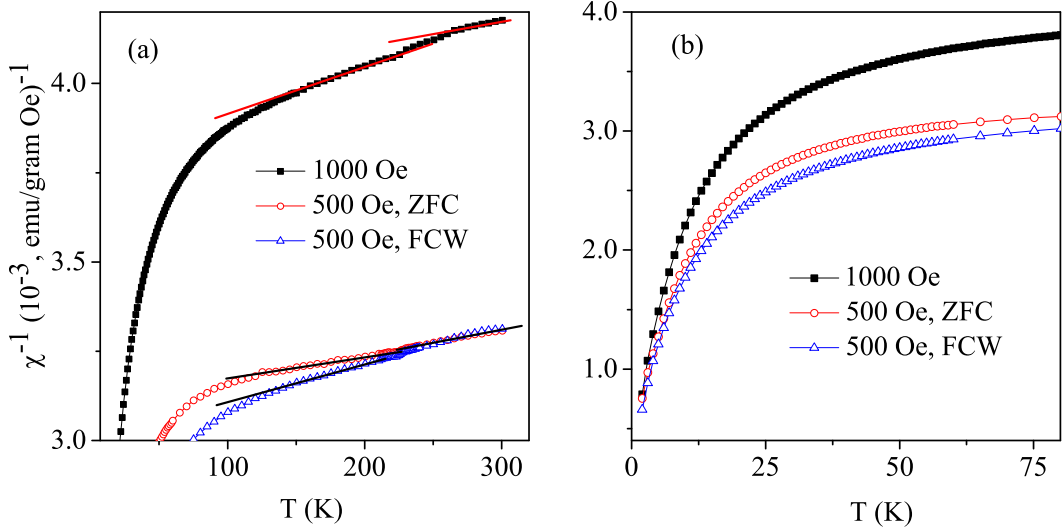}
\caption{ (a) Inverse susceptibility ($\chi^{-1}$)  versus temperature ($T$) plots of  1$T$-TaS$_2$ at high temperatures. It shows a change in the slope of $\chi^{-1}$ which indicates the transition between nearly commensurate and commensurate CDW states. (b) The low temperature susceptibility of 1$T$-TaS$_2$, which shows downward curvature and therefore does not follow the Curie-Weiss law expected in a system with fixed number of free spins. Linear solid lines are guide to the eye.}
\end{figure}

$Region$ $(i)$: In Fig. {\color{blue}3(a)}, we present the $\chi^{-1}$ versus $T$ response of  1$T$-TaS$_2$ obtained with ZFC protocol at $H$ = 1000 Oe and with ZFC and FCW protocols at $H$ = 500 Oe above $T$ = 75 K. In this range, as we reduce temperature from $T$ = 300 K, the susceptibility monotonically increases and the rise is nearly linear. As a result, inverse susceptibility decreases linearly. In this temperature range, 1$T$-TaS$_2$ resides in the nearly commensurate CDW phase. In the nearly commensurate phase, the CDW is rotated away from the lattice. The star-of-David clusters start forming in this phase and they are arranged in the form of hexagonal domains. The CDW phase is commensurate in a domain separated by diffuse discommensurations {\color{blue}\cite{Thomson1994, Wang2019}}. At $H$ =  1000 Oe, there is a noticeable change in the slope in $\chi$ versus $T$ data around $230< T< 265$ K. This is more prominent in the temperature dependence of inverse susceptibility as shown in Fig. {\color{blue}3(a)} (black filled square). The change in the slope can also be observed in the ZFC and FCW curves measured at $H$ = 500 Oe. It is interesting to note that the bifurcation between the ZFC and FCW curves at $H$ = 500 Oe also starts below this temperature. This temperature region tallies with the temperature region over which the first order transition between the commensurate to nearly commensurate CDW phases takes place {\color{blue}\cite{Thomson1994, Kratochvilova2017,Wang2019}}. It may be noted that this transition between the commensurate to nearly commensurate CDW is more clearly observed in temperature variation of resistivity {\color{blue}\cite{Yu2015}}. There, it shows pronounced thermal hysteresis between cooling and heating cycles which indicates the first order nature of the transition. However, in magnetization measurements, this transition exhibits prominent step like anomaly only at high magnetic fields {\color{blue}\cite{Dai1995}}. 

It may be noted that, another CDW state called as triclininc phase has been reported as stable phase in this temperature range which is only observed during heating {\color{blue}\cite{Thomson1994, Wang2019}}. The scanning tunneling microscopy studies have revealed that in this phase 1$T$-TaS$_2$ loses the hexagonal domain structure of star-of-David clusters and instead forms long and narrow stripes {\color{blue}\cite{Thomson1994, Wang2019}}. The formation of star-of-David indicates that even in these CDW phases, local moments should exist, but the number and the arrangement of the moments should depend on the detail domain structure. Now, if in this temperature region, exchange interaction between the local moments is not dominating, then the moments behaves like free spins. In that case, the increase in the susceptibility as observed in Fig. {\color{blue}2(a)} might be an indication that as the temperature decreases, the number of the free spins also increases. We tried to fit the susceptibility above $T$ = 250 K by using Curie-Weiss law, but it yields unphysical parameters.  

$Region$ $(ii)$: Below $T$ = 250 K, the susceptibility increases relatively fast. In this temperature range, 1$T$-TaS$_2$ exhibits various interesting properties. To begin with, the CDW phase is commensurate with the underlying lattice and forms a hexagonal lattice with a single star-of-David cluster sitting at each site with $\sqrt{13}\times \sqrt{13}$ reconstructed unit cell {\color{blue}\cite{LawLee2017}}. Secondly, in each star-of-David, we have one unpaired electron. Based on the electron band theory, it should be a metal, but electron-electron correlation localizes the electrons and induce transition from metal to Mott insulating state, with a local moment S = 1/2 per star-of-David cluster arranged on a homogeneous triangular lattice. Thirdly, as also mentioned earlier, the bifurcation between the ZFC and FCW curves starts as observed at $H$ = 500 Oe.

\begin{figure}[t]
\centering
\includegraphics[scale=0.38]{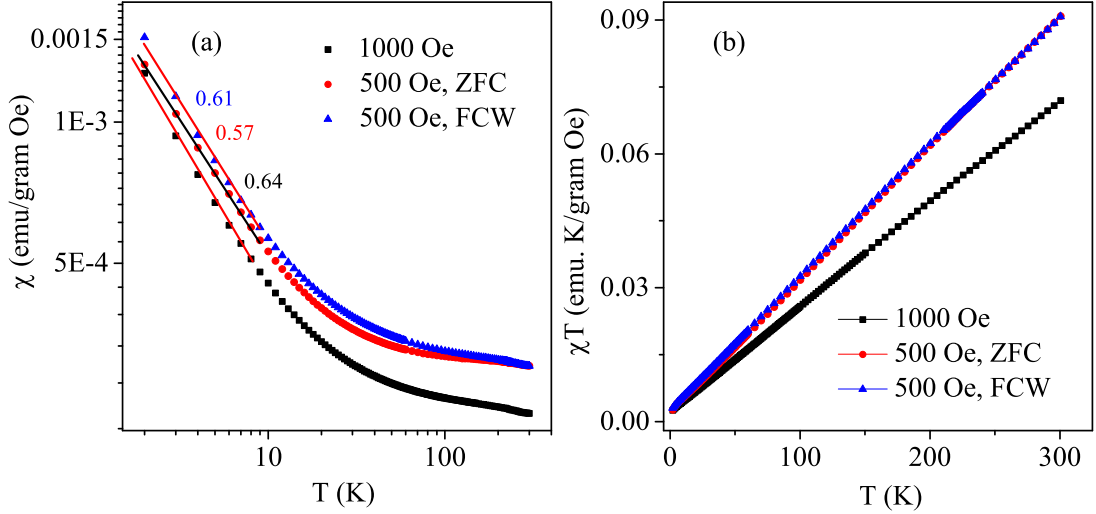}
\caption{ (a) log $\chi$ versus log $T$ plot of  1$T$-TaS$_2$ obtained under ZFC protocol at H = 1000 and 500 Oe and (b) temperature dependence of $\chi T$ with T at $H$ = 1000 Oe and the ZFC and FCW curves at $H$ = 500 Oe. Solid red lines in (a) below T = 10 K are the linear fit. The slope of the straight lines give the value of $\alpha$.}
\end{figure}
In $region$ $(iii)$, susceptibility increases rapidly with decrease in the temperature down to the lowest measured temperature of 2 K. In Fig. {\color{blue}3(b)}, we have shown the temperature dependent $\chi^{-1}$ at low temperatures, which clearly exhibits the deviation from linearity.  A Mott insulator is generally expected to have an antiferromagnetic spin arrangement because the system gains energy of $J = 4t^2/U$ by virtual hopping between neighboring sites when these spins are oppositely aligned. Here $J$ is the magnetic exchange interaction between the two nearest neighbor spins, $t$ is the amplitude for the electron to hop from one lattice site to the other, and $U$ is the Coulomb interaction energy between two electrons with different spin directions. However, 1$T$ -TaS$_2$ does not reveal any long range magnetic order. Note that, the inverse susceptibility shows strong non-linearity in this temperature region and deviates downward from the Curie-Weiss like paramagnetic response of the spins given by $\chi = \frac{C}{T-T_0}$.  

In 1$T$-TaS$_2$, the regular arrangement of star-of-David clusters in the commensurate phase gives rise to S = $\frac{1}{2}$ arranged on perfect 2D-triangular lattice, which is similar to organic QSL compounds, such as  $\kappa$-(BEDT-TTF)$_2$Cu$_2$(CN)$_3$, EtMe$_3$Sb[Pd(dmit)$_2$]$_2$ etc. These organic materials have antiferromagnetic interactions of the order of $J/k_B$ = 200 K and the magnetic susceptibility shows entirely different temperature dependence as compared to 1$T$-TaS$_2$ {\color{blue}\cite{Tamura2002}}. It initially increases as the temperature is reduced from $T$ = 300 K, shows a broad hump at some intermediate temperature depending upon the strength of the exchange interaction and decreases again down to lowest temperature {\color{blue}\cite{Tamura2002}}. In 1$T$-TaS$_2$, the nature and the strength of the exchange interaction between localized moments is not yet known. Earlier, the low temperature susceptibility below $T$ = 150 K has been analyzed using modified CW law given by $\chi = \chi_0 + \frac {C}{T-T_0}$ in a number of works {\color{blue}\cite{DiSalvo1980,Ribak2017,Guy1982,Furukawa1984}}. Unanimously, the strength of the interaction has been found to be very small. However, the magnitude, as well as the sign of the exchange interaction vary with sample preparation condition. For instance, Ribak et al has reported a weak antiferromagnetic interaction of the order of $J/k_B \approx$ 1 K, which is very small compared to other spin liquid candidates. 

It may be noted here that similar sharp rise in magnetic susceptibility has been observed in doped semiconductors, such as phosphorous, boron doped Si {\color{blue}\cite{Andres1981,Sarachik1985, Sarachik1986}}. These systems undergo a metal to insulator transition below a critical concentration of doping. For example, in Si:P system, it shows insulating behavior below $n_d = 1\times 10^{18}$$cm^3$, which implies the donor electrons are localized and give rise to local magnetic moments. The exchange interaction between spins are random in magnitude and antiferromagnetic in nature {\color{blue}\cite{Andres1981,Sarachik1985, Sarachik1986}}. Importantly, these systems are more like amorphous antiferromagnet and show distinct behavior than spin glasses and rather behaves like random singlet state. Spin susceptibility shows rapid rise at low temperature and deviates from CW law. It rather follows a power law dependence ($T^{-\alpha}$, $\alpha <1$) and the $log$ $\chi$ versus $log$ $T$ curve exhibits linear behavior within certain range of temperatures, where the slope of the straight line gives the value of the exponent $\alpha$.  We have also plotted $log$ $\chi$ versus $log$ $T$ data of the ZFC curve of 1$T$-TaS$_2$ at H = 1000 Oe, and the ZFC and FCW curves at $H$ = 500 Oe  in Fig. {\color{blue}4(a)}. Note that, in the temperature range of 2-10 K it shows good linear behavior. However, the value of $\alpha$, which has been obtained from slope of the fitted linear curve below 10 K changes depending on the measurement field and protocols. At $H$ = 1000 Oe,  a slope of $\alpha$ = 0.64 has been found. On the other hand, at $H$ = 500 Oe, the values of $\alpha$ are 0.61 and 0.57 for FCW and ZFC curves respectively. Similar values of the exponent have also been reported in Si:P system in wide range of doping concentrations and a few other spin liquid candidates {\color{blue}\cite{Sarachik1985, Sarachik1986, Volkov2020}}. The dependence of the exponent $\alpha$ on the applied magnetic field as well as the measurement protocol is interesting.  Also note that, the change in slope around $T$ = 230 K, which we have discussed earlier (see Fig. {\color{blue}3(a)}) is more pronounced in $log$ $\chi$ versus $log$ $T$ data in Fig. {\color{blue}4(a)}.

In Fig. {\color{blue}4(b)}, we have further plotted the $\chi T$ versus $T$ at $H$ = 1000 and 500 Oe. We note here that in systems, where the number and magnetic moment of the localized spins are unambiguously known, the temperature dependence of $\chi T/C$ (where $C= N\mu^2/3k_B$) gives important information about the exchange interaction {\color{blue}\cite{Vobo2018}}. For paramagnetic system, $\chi T/C = 1$, and $\chi T/C> 1$, $\chi T/C < 1$ for ferromagnetic and antiferromagnetic systems respectively. In case of 1$T$-TaS$_2$, the contribution of the magnetic moment from localized electrons, i.e. the value of C is not known. Therefore, we have only plotted the $\chi T$ instead of $\chi T/C$ and surprisingly the $\chi T$ is linear in the entire temperature range of 2-300 K at both $H$ = 1000 and 500 Oe, and the slope of the linear curve depends on the measurement field. In random singlet state, two spins at random distances form singlet pairs and there can be orphan spins, which are not paired to other spins. How these unpaired spins behave as a function of  temperature and field is not yet known. On the other hand, in 1$T$-TaS$_2$, as temperature decreases, the possibility of change in the number of unpaired electrons which originates from cluster of David star and gives rise to local spin moment can not be disregarded. It will make the value of $C$ itself temperature and field dependent.

\section{Summary}
Summarizing, we have studied in details the temperature dependence of magnetic susceptibility of the quantum spin liquid compound 1$T$-TaS$_2$ at low magnetic fields. Prima facie susceptibility appears to be temperature independent down to nearly T = 50 K, followed by a sharp rise at further lower temperatures. However, a closer look reveals that a small but significant temperature dependence exists even in the nearly commensurate CDW state, which implies the formation of magnetic moments much above the Mott transition temperature. Inverse susceptibility is nearly linear above $T$ = 200 K and exhibits a noticeable change in the slope around $230< T< 265$ K, which can be correlated to the process of first order phase transition between nearly commensurate and commensurate CDW phases. At low measurement fields, the ZFC and FC curves also start to bifurcate around the same temperature range. At further lower temperatures, particularly below $T$ = 50 K, the susceptibility increases very rapidly down to lowest temperature. The observed behaviour cannot be not explained well with the help of a modified Curie-Weiss (CW) law given by: $\chi = \chi_0 + \frac {C}{(T-T_0)}$. There is a discernable deviation from the CW behaviour. Below $T$ = 10 K, the susceptibility follows a power law, $T^{-\alpha}$ expected from random singlet state, where the value of $\alpha$ has been found to depend on the measurement field and the measurement protocols.   

\section{Acknowledgment}
S. B. Roy acknowledges financial support from Department of Atomic Energy, India in the form of Raja Ramanna Fellowship

\end{document}